\documentclass[10pt]{article}
\usepackage{latexsym}
\usepackage{amssymb}
\usepackage{amsmath}
\usepackage{amscd}
\usepackage{amsthm}
\usepackage[left=1.5cm,top=1.5cm,right=1.5cm,bottom=1.5cm]{geometry}
\usepackage{graphicx}
\usepackage{textcomp}
\usepackage{hyperref}
\usepackage[font={footnotesize,it}]{caption}

\begin{document}
\begin{center}
\large{\bf{PROBING DARK ENERGY IN THE SCOPE OF BIANCHI TYPE I SPACETIME}} \\
\vspace{10mm}
\normalsize{Hassan Amirhashchi}\\
\vspace{5mm}
\normalsize{Department of Physics, Mahshahr Branch, Islamic Azad University,  Mahshahr, Iran \\
E-mail:h.amirhashchi@mhriau.ac.ir},~~~ hashchi@yahoo.com \\
\end{center}
\begin{abstract}
It is well known that flat FRW metric is a special case of Bianchi type I spacetime.
In this paper, we use 38 Hubble parameter, $H(z)$, measurements at intermediate redshifts $0.07\leq z\leq 2.36$ 
and its joint combination with the latest \textgravedbl joint light curves\textacutedbl (JLA) sample, comprised of 740 type Ia supernovae in the
redshift range $z \epsilon [0.01, 1.30]$ to constrain the parameters of Bianchi type I dark energy model. We also use the same datasets to constrain
flat $\Lambda$CDM Model. In both cases, we specifically address the expansion rate $H_{0}$ as well as the transition redshift $z_{t}$
determinations out of these measurements. In both Models we found that using joint combination of datasets gives rise to lower values
for model parameters. Also to compare the considered cosmologies, we have made Akaike information criterion (AIC) and Bayes factor ($\Psi$) tests.
 \end{abstract}
\section{INTRODUCTION}
Using supernovae type Ia (SNe Ia) as standard candles enabled us to probe the history of cosmic expansion at high accuracy i.e at low
and intermediate redshifts. Observations from Various independent research teams show that the expansion of our universe is speeding up
at present time \cite{ref1}-\cite{ref5}. This phase transition from decelerating to accelerating expansion is due to an unknown mechanism
changing the sign of the universal deceleration parameter $q(z)$. In spite of considerable efforts, still the correct physical
explanation of such transition is one of the must challenging fields in cosmology. However, it is commonly accepted that the current
cosmic acceleration could be investigated (1) by assuming an unknown and unusual source of energy with negative
pressure called dark energy (DE) in cosmic fluid or (2) by assuming the general theory of gravity is modified.
Within the framework of General Relativity (GR), the simplest way for explaining such a mechanism is by assuming a cosmological constant
$\Lambda$ \cite{ref6,ref7} in Einstein field equations. However, this scenario suffers from cosmological and coincidence problems \cite{ref8,ref9,ref10}.
Exploring of DE is possible through its equation of state parameter (EoS) connecting pressure and density of DE
as $p=\omega^{X} \rho$ (where $p$ is the pressure and $\rho$ is the density of DE) as well as its microphysics, characterized
by the speed of sound $c_{s}^{2}$. It is worth mentioning that the study of a fluid model of dark energy requires considering both
an equation of state parameter (EoS) and sound speed $c_{s}^{2}$ at the same time. By the way,
as mentioned by Sandage \cite{ref11} in both static and dynamical cases, the basic characteristics of the cosmological evolution could be expressed
in terms of the Habble ($H_{0}$) and deceleration ($q_{0}$) parameters which enable us to construct model-independent
kinematics of the cosmological expansion.\\

Long time (since 1929) measurements of Hubble constant, $H_{0}$, indicated that the value of this parameter to be between
50-100 km s$^{-1}/$Mpc  \cite{ref12,ref13}. Improving the control of systematics, use of different and novel calibration techniques and by the aid of
space facilities, now we can estimate the observed value of $H_{0}$ more accurately. Comparing different observations techniques
indicate that the indirect estimates of the Hubble constant lead to lower values of $H_{0}$ compared with direct estimations.
For example, the most recent direct estimate of $H_{0}$ obtained by Riess et al.\cite{ref14} is $H_{0}=(73.0 \pm 1.8)$km s$^{-1}/$Mpc while
the indirect measurement of the expansion rate of our universe such as the observations of the cosmic microwave back-
ground (CMB) anisotropy by WMAP \cite{ref15}, Planck Collaboration \cite{ref16} and the joint Atacama Cosmology Telescope (ACT) \cite{ref17}
and the WMAP 7-year cosmic microwave background (CMB) anisotropy data \cite{ref18} yielded $H_{0}=(70.0 \pm 2.2)$km s$^{-1}/$Mpc
, $H_{0}=(67.27 \pm 0.66)$km s$^{-1}/$Mpc and $H_{0}=(70.0 \pm 2.4)$km s$^{-1}/$Mpc respectively. The best technique for deriving a
convincing summary observational estimate of $H_{0}$ is the median statistics technique \cite{ref19}-\cite{ref25}. Applying median
statistics to 331 $H_{0}$ estimates listed by Huchra \footnote{https://www.cfa.harvard.edu/∼dfabricant/huchra/}, Gott et al \cite{ref19}
obtained $H_{0}=(67 \pm 3.5)$km s$^{-1}/$Mpc. From 461 measurements (up to the middle of 2003) Chen et al \cite{ref21}
found $H_{0}=(68 \pm 3.5)$km s$^{-1}/$Mpc, and from 553 measurements (up to early 2011) Chen \& Ratra \cite{ref26,ref27}
found $H_{0}=(68 \pm 2.8)$km s$^{-1}/$Mpc. It is worth noting that since the expansion rate of the universe
,$H_{0}$, is degenerate with the SN absolute magnitude, the observations of SN Ia by themselves could not 
evaluate a value for the local expansion rate of the Universe. However, as proposed by Jimenez
et al \cite{ref28}, passively evolving red galaxies whose age can be precisely estimated from a spectroscopic analysis, could be used
to provide the redshift dependence of the cosmic expansion rate $H(z)$. Liu et al \cite{ref29} have recently used these observational
Hubble data (OHD) and found a value of $H_{0}=(68.6)$km s$^{-1}/$Mpc. It should be emphasized that a more carefulness has to be taken when
one uses several SN datasets such as Union as these datasets provide cosmological distance moduli that are derived assuming
a flat $\Lambda$CDM model. Hence, to constrain cosmological models that are different from $\Lambda$CDM we
can use \textgravedbl latest joint light curves\textacutedbl (JLA) dataset\footnote{All data used are available on
http://supernovae.in2p3.fr/sdss-snls-jla/ReadMe.html} which provides model-independent apparent magnitudes instead of
model-dependent distance moduli. Recently, Lukovic et al \cite{ref30} have used JLA dataset to constrain different class of models
based on Lema\^{i}tre-Tolman-Bondi (LTB) metric. Reader is advised to see Refs\cite{ref31,ref32,ref33,ref34,ref35} for more detailed study of LTB space-time.
In this paper we consider $38 H(z)$ data points at the redshift range $0.07<z<2.36$ compiled by Farooq and Ratra \cite{ref36}, JLA dataset
and their combination to constrain different models based on Bianchi type I (henceforth BI) metric which describes a homogeneous but
anisotropic Universe.\\
The plan of the paper is as follows. Sec. 2 deals with the theoretical models we considered. In Sec 3 we briefly summarize
the data analysis method we use. In Subsec 3.1 we constrain $\Lambda$CDM model. We constrain $\omega$BI model in Subsec 3.2.
We derive the transition redshift for both models in Subsec 3.3. Finally, In Sec. 4 we summarize our findings and conclusions.
\section{THEORETICAL MODELS}
The assumption of considering space-time to be homogeneous and isotropic (cosmological principle) could be suffers due to the following two
facts. First, is the recent cosmic observations which indict small variations between the intensities of the microwaves coming
from different directions in the sky. Second, is the fact that at very large scales (beyond event horizon) and also at the small scales,
universe necessarily does not have the same symmetries as we consider for FRW spacetime. Therefore, it will be more reasonable
to obtain \textgravedbl almost FRW\textacutedbl models representing a universe that is \textgravedbl FRW-like\textacutedbl on
large scales but allowing for generic in inhomogeneities and anisotropies arising during structure formation on a small scale. To be
able to compare detailed observations, we take Bianchi type I (BI) metric which is homogeneous but anisotropic. It is shown by
Goliath and Ellis \cite{ref37} that some Bianchi models isotropise due to inflation.\\
In an orthogonal form, the Bianchi type I line-element is given by
\begin{equation}
\label{eq1}
ds^{2} = -dt^{2} + A^{2}(t)dx^{2}+B^{2}(t)dy^{2}+C^{2}(t)dz^{2},
\end{equation}
where $A(t), B(t)$ and $C(t)$ are functions of time only. \\

The Einstein's field equations ( in gravitational units $8\pi G = c = 1 $) read as
\begin{equation}
\label{eq2} R^{\mu}_{\nu} - \frac{1}{2} R g^{\mu}_{\nu} = T^{(m) \mu}_{\nu} +
T^{(X) \mu}_{\nu},
\end{equation}
where $T^{(m) \mu}_{\nu}$ and $T^{(X) \mu}_{\nu}$ are the energy momentum tensors of dark matter and dark energy,
respectively. These are given by
\[
  T^{m \mu}_{\nu} = \mbox{diag}[-\rho^{m}, p^{m}, p^{m}, p^{m}],
\]
\begin{equation}
\label{eq3} ~ ~ ~ ~ ~ ~ ~ ~  = \mbox{diag}[-1, \omega^{m}, \omega^{m}, \omega^{m}]\rho^{m},
\end{equation}
and
\[
 T^{X \mu}_{\nu} = \mbox{diag}[-\rho^{X}, p^{X}, p^{X}, p^{X}],
\]
\begin{equation}
\label{eq4} ~ ~ ~ ~ ~ ~ ~ ~ ~ ~ ~ ~ ~ ~ = \mbox{diag}[-1, \omega^{X}, \omega^{X},
\omega^{X}]\rho^{X},
\end{equation}

where $(\rho^{m}, p^{m} )$ and $(\rho^{X}, p^{X} )$ are the energy density and pressure of the dark matter (DM) and dark energy (DE) respectively. 
Energy density and pressure of each component are related by the equation of state (EoS) $p=\omega \rho$.
The 4-velocity vector $u^{i} = (1, 0, 0, 0)$ is assumed to satisfy $u^{i}u_{j} = -1$.
As shown in Refs \cite{ref38,ref39} the exact solution of the Einsteins field equations in a co-moving coordinate system ($u^{i} = \delta^{i}_{0}$) leads to
the following {\bf Friedmann-Like Equations}
\begin{equation}
\label{eq5} \left(\frac{\dot{a}}{a}\right)^{2}=\frac{1}{3}\sum_{i}\rho_{i}+Ka^{-6},~~~~K<0
\end{equation}
\begin{equation}
\label{eq6} \frac{\ddot{a}}{a} = -\frac{1}{6}\sum_{i}\rho_{i}(1+3\omega_{i}),
\end{equation}
where $a=(ABC)^{\frac{1}{3}}$ is the average scale factor. In term of anisotropy parameter $A_{m}$, eq (\ref{eq5}) could be written as
(see Appendix A)
\begin{equation}
\label{eq7} \left(\frac{\dot{a}}{a}\right)^{2}=\frac{1}{3}\sum_{i}\rho_{i}+\frac{A_{m}}{6},
\end{equation}
It is worth to mention that the constant $K$ also is a scale which denotes the
deviation from isotropy, therefore $K=0$ (and equivalently $A_{m}=0$) represents flat FRW universe. From eq (\ref{eq6}) it is clear that
to get accelerating expansion (i.e $\ddot{a}>0$) at least one of the components of the cosmic fluid must have $\omega<-\frac{1}{3}$.
Using eq (\ref{eq6}) one can find the dimensional Hubble parameter as
\begin{equation}
\label{eq8} E(z)=\frac{H}{H_{0}}=\sqrt{\sum_{i}\Omega_{i}(1+z)^{3(1+\omega_{i})}+\tilde{A}_{m}(1+z)^{6}},
\end{equation}
where $H(0)=H(z=0)$, $\tilde{A}_{m}=(\frac{A_{m0}}{H_{0}^{2}})$ and $\Omega_{i}=3H_{0}^{2}\rho_{i}$ is the current density of the i-th component.
Also the law of energy-conservation equation ($T^{\mu\nu}_{;\nu} = 0$) yields
\begin{equation}
\label{eq9} \sum_{i}\dot{\rho_{i}}+3(1+\omega_{i})\rho_{i}H = 0,
\end{equation}
which shows the density evolution of each component of the cosmic fluid. While we consider various EoS parameter
for the DE models, the DM usually considered cold with zero EoS parameter ($\omega^{m}=0$). In our study we
discuss the observational constraints on the free parameters of the following models.\\

{\bf Model I}. We consider flat $\Lambda$CDM with $\omega^{X}=-1$ which could be obtained by putting $K=0$ in eq (\ref{eq6}).
This is the simplest way of best-fitting current cosmological observations. The base parameters set for the Model I is
\[
 {\bf P}=\{\Omega^{X\equiv \Lambda},\Omega^{m}, n_{s}, H_{0}, Age/Gyr, \sigma_{8}\}.
\]

{\bf Model II}. We also consider a $\omega$BI model with a EoS parameter which is constant with respect to the cosmic expansion.
In this case the best set of parameters is

\[
 {\bf P}=\{\Omega^{X},\Omega^{m}, n_{s}, H_{0}, Age/Gyr, \sigma_{8}, A_{m}, \omega^{X}\}.
\]\\

\section{DATA SETS AND RESULTS}
The main goal of this paper is to constrain the free parameters of the theoretical models described above using independent
observables that are SN Ia (JLA dataset), OHD and their joint combination which could increase the sensitivity of our estimates.
We specifically address the $H_{0}$ determination out of these measurements. In this section we briefly describe the
observational data sets used in this work.\\

\vspace{2mm}
\noindent
{\bf SN Ia:}
As mentioned above the most advantage of the (JLA)‍‍‍ dataset is the model-independent property of this dataset.
The JLA dataset we consider in this paper is comprised of $740$ type Ia supernovae in the redshift range $0.01\leq z\leq 1.30$ \cite{ref40}.
To calculate the corresponding likelihood (${\it L_{SN}}$) and hence constrain the models parameters, we follow the method
given by Tr\o{}st Nielsen et al \cite{ref41}. Note that, In analysing the JLA dataset by itself, we do not consider any prior on $H_{0}$.
As we will show bellow, although the JLA dataset by itself is not sensitive to $H_{0}$, However,
in the joint analysis, JLA affect the estimate of $H_{0}$.\\

\vspace{2mm}
\noindent
{\bf OHD:} Farooq and Ratra have recently compiled a $38H(z)$ datapoints in the redshift range $.07\leq z\leq 2.36$ \cite{ref36}.
To constrain cosmological parameters, we use this dataset and maximize the following likelihood.
\begin{equation}
\label{eq10} {\it L_{OHD}}\propto exp\left[-\frac{1}{2}\sum_{i=1}^{38}\left(\frac{H^{th}(z_{i},{\bf P})-H^{obs}(z_{i})}{\sigma^{2}_{H,i}}\right)^{2}\right],
\end{equation}
where $H^{th}$ is the predicted value of $H(z)$ in the cosmological model given by eq (\ref{eq7}), $H^{obs}(z_{i} )$ is the measured value with
variance $\sigma^{2}_{H,i}$ at redshift $z_{i}$ and {\bf P} represents the free parameters of the cosmological model.\\

\vspace{2mm}
\noindent
{\bf Combined Analysis:} In what follows we show that both {\bf JLA} and {\bf OHD} datasets 
provide good estimates of the cosmological parameters. However, to obtain more stringent constraints,
we combine both datasets. One of the advantage of combining datasets is that the joint analysis allows us to provide
separate estimates for the absolute magnitude of SN and the Hubble constant $H(z)$. To evaluate the total likelihood as the
product of the likelihoods of the single datasets, we assume that the datasets are independent. The total likelihood is given by
\begin{equation}
\label{eq11} {\it L_{tot}= L_{SN}L_{OHD}}.
\end{equation}
To obtain correlated Markov Chain Monte Carlo (MCMC) of the samples, we modify CLASS \cite{ref42}
and Monte Python \cite{ref43} codes and use Metropolis Hastings algorithm with uniform priors on the model
parameters. To analyze the MCMC chains we use the GetDist Python package \cite{ref44}.

\subsection{CONSTRAINTS ON FLAT $\Lambda$CDM MODEL}
The flat $\Lambda$CDM model which is commonly considered as the concordance model in cosmology is the simplest model
for the best-fitting of current observations. Therefore, this model has been tested with almost all the available
cosmological observables. In this case, the results of our statistical analysis are shown in Table 1. The contour plots of the Model I
parameters are also depicted in Figure 1. From Table 1 we observe that our values for $\Omega^{m}$ (at $1\sigma$ error), obtained
from OHD and OHD+JLA are in high agreement from the recent determination of Planck: TT, TE, EE + lowP, which
provided $\Omega^{m}= 0.316 \pm 0.009$.\\
\begin{figure}[htb]
\centering
\includegraphics[width=15cm,height=15cm,angle=0]{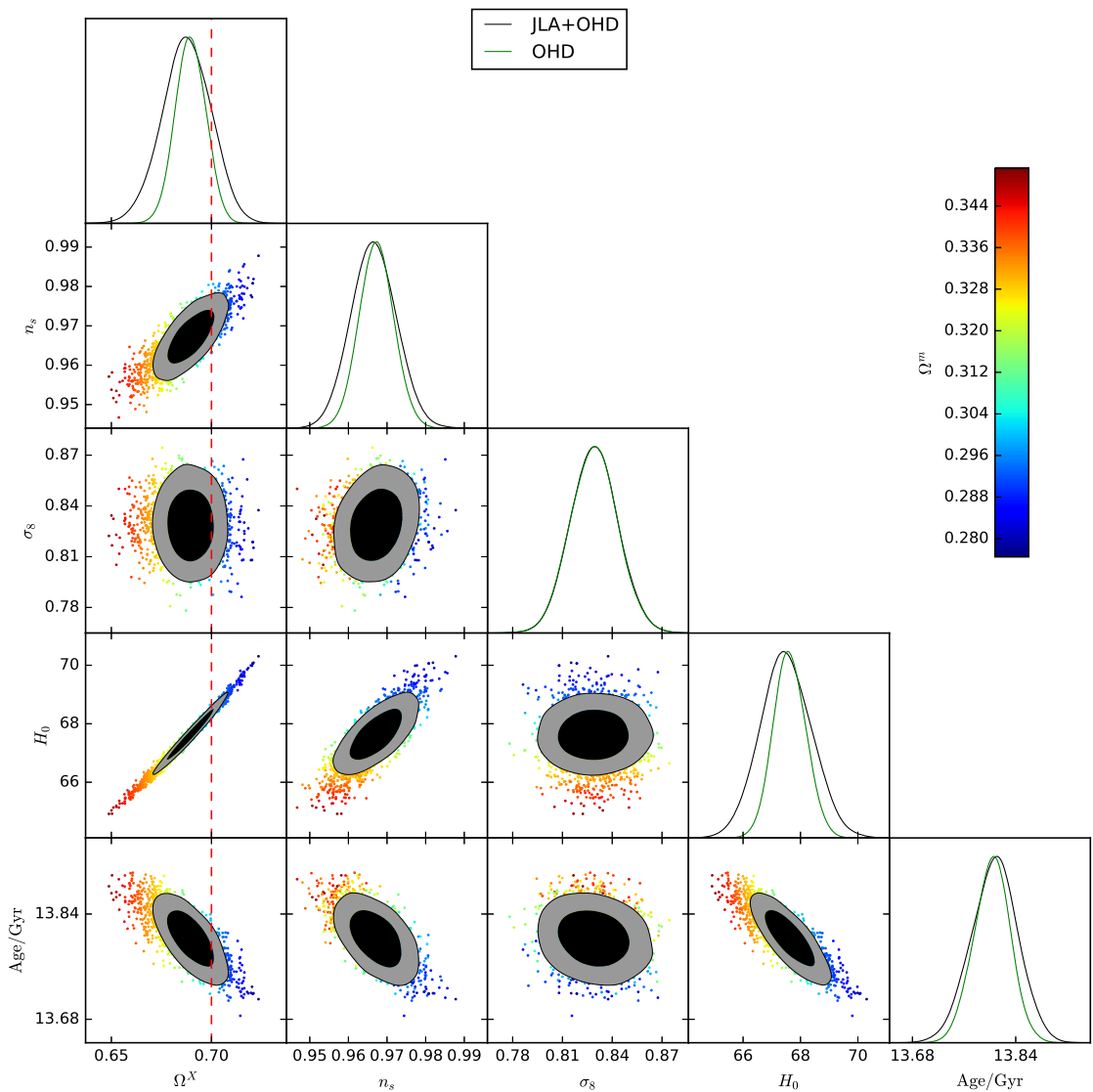}
\caption{One-dimensional marginalized distribution, and two-dimensional contours with $68\%$ CL and $95\%$ CL for the model parameters.
The vertical dotted Red line stands for $\Omega^{\Lambda}=0.7$.}
\end{figure}

\begin{table}[ht]
\caption{Results from the fits of the $\Lambda$CDM model to the data at 1$\sigma$ confidence level.}
\centering
\begin{tabular} {ccccccccc}
\hline
Parameter    & JLA & OHD & JLA+OHD  \\[0.5ex]
\hline
\hline{\smallskip}
$H_{0}$ &  - &    $68.3 \pm 1.5$ &  $67.9 \pm 1.2$\\ 

$\Omega^{\Lambda}$ &  $0.624\pm 0.031$   & $0.6890\pm 0.0076$ &  $0.687\pm 0.012$\\

$\Omega^{m}$ & $0.376\pm 0.031$  &  $0.3110\pm 0.0076$  & $0.313\pm 0.012$ \\ 

$n_{s}$ & $0.959\pm 0.019$ &  $0.9671\pm 0.0043$ & $0.9662\pm 0.0057$ \\

$\sigma_{8}$ & $0.835\pm 0.042$ & $0.828\pm 0.015$ & $0.829\pm 0.015$ \\

$Age/Gyr$ & $13.808^{+0.039}_{-0.035}$ & $13.794\pm 0.046$ & $13.803^{+0.031}_{-0.027}$\\[0.5ex]

\hline
\end{tabular}
\label{table:nonlin}
\end{table}
Moreover, our estimates for the current expansion rate, $H_{0}$ derived from OHD alone and OHD+JLA  
are in good agreement with those obtained by chen \& Ratra ($68\pm 2.8$) \cite{ref26}, Ade et al,Planck, ($67.8\pm 0.9$) \cite{ref45}
and Aubourg et al,BAO, ($67.3\pm 1.1$) \cite{ref46}. It is worth mentioning that although the final value for $H_{0}$ derived from the
joint JLA+OHD analysis is in high agreement with the Planck (2015) and BAO (see Figure 2), but its difference
from the findings by Efstathiou \cite{ref47} and Riess et al \cite{ref48} is by $1.9\sigma$ and $2.6\sigma$ respectively.
It is important to note that the JLA dataset by itself is not sensitive to the expansion rate $H_{0}$, but when we combine
it to the other datasets, in the joint analysis, JLA constrains other parameters of the Model under consideration, which in turn
affect the estimate of $H_{0}$. The robustness of our fits can be viewed by looking at figures 2 \& 3. From this figure we
observe that the joint analysis give rises to a better fits to the observational data. \\

\begin{figure}[htb]
\begin{minipage}[b]{0.5\linewidth}
\centering
\includegraphics[width=8cm,height=6cm,angle=0]{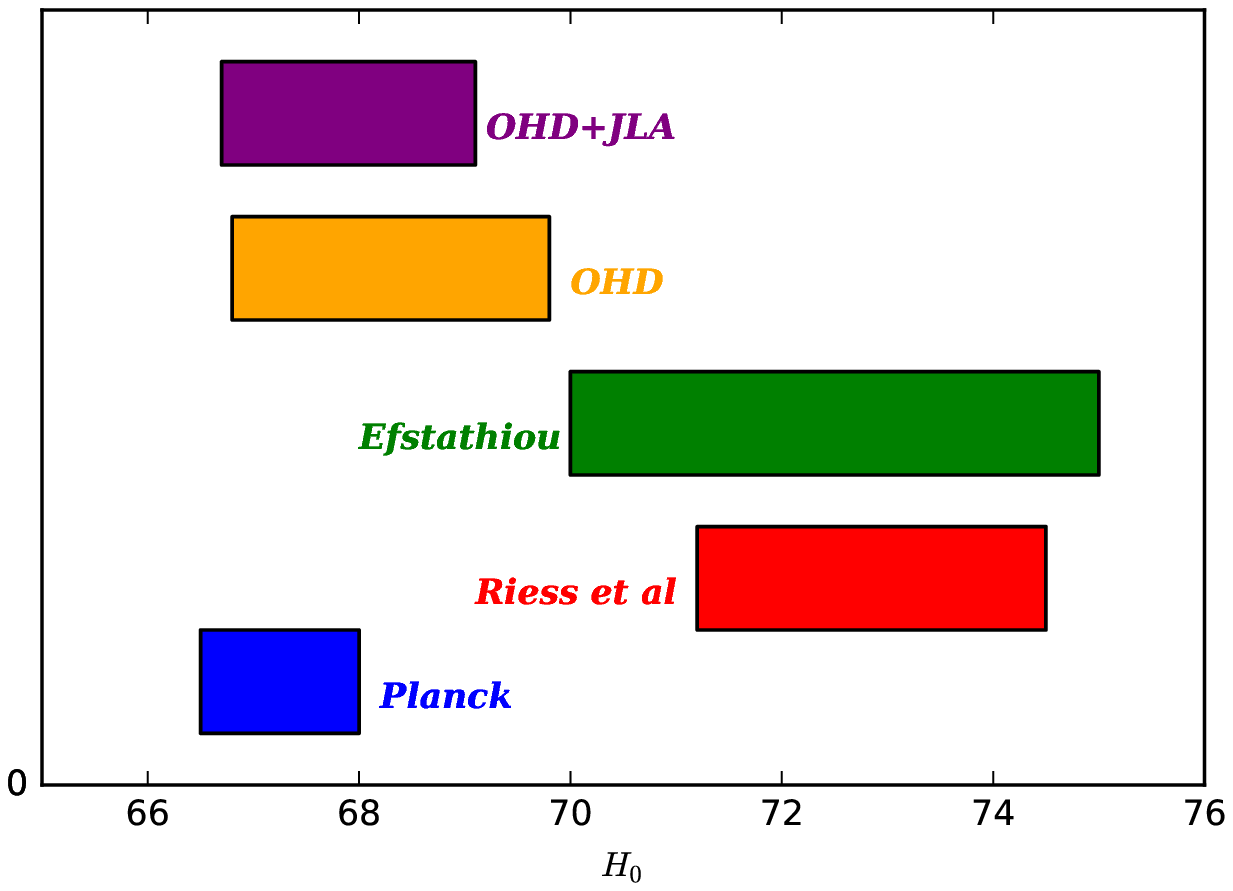} \\
\caption{A schematic representation at $1\sigma$ CL of $H_{0}$ for the Model I when fitted to OHD (gold color) and 
two datasets combined (purple). Constraints from the direct measurement by Riess et al. (2016) (red color), the reanalysis by Efstathiou
(2014) (green color), and the Planck Collaboration (2015) (blue) are also shown.}
\end{minipage}
\hspace{0.5cm}
\begin{minipage}[b]{0.5\linewidth}
\centering
\includegraphics[width=8cm,height=6cm,angle=0]{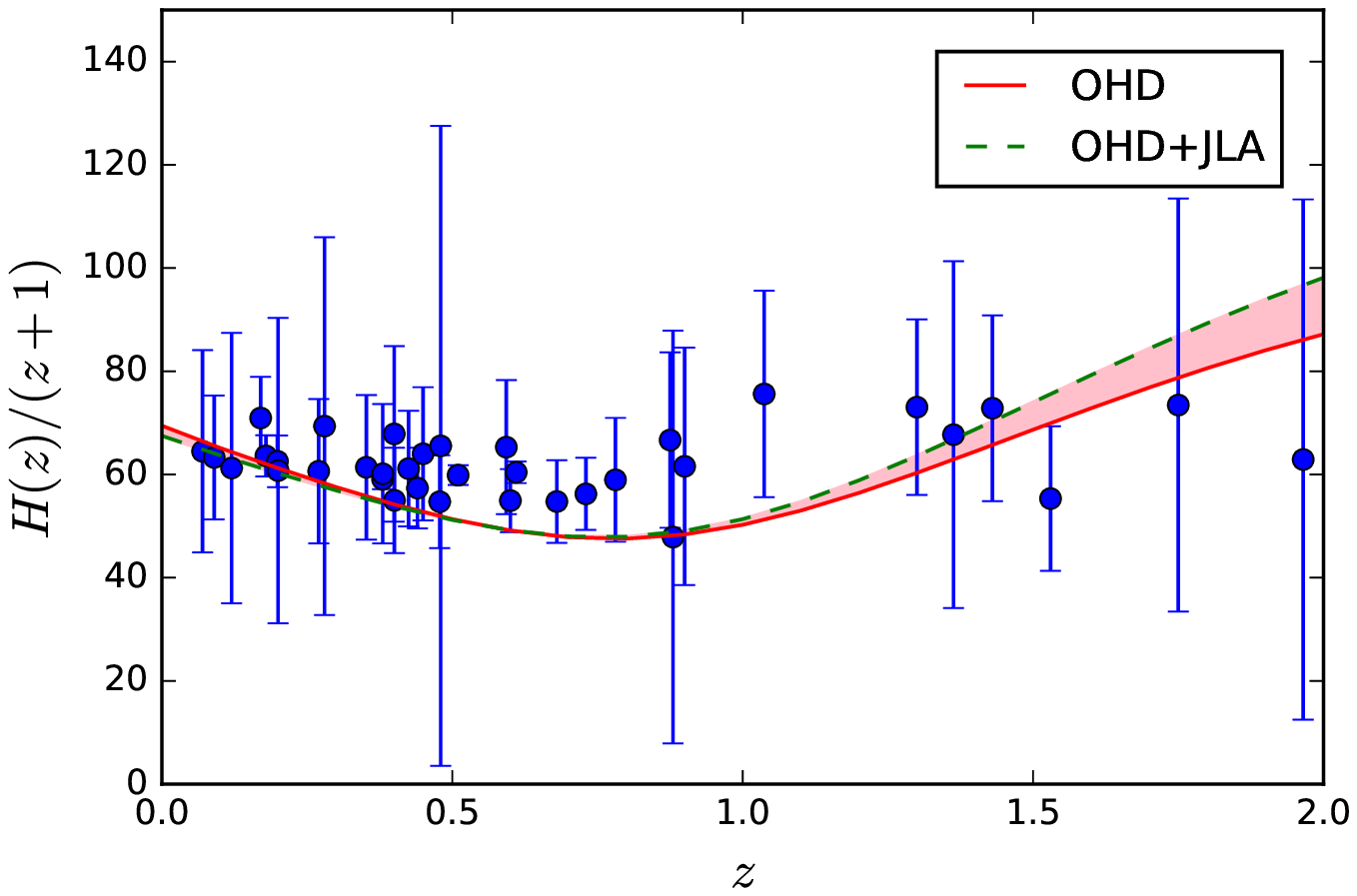}
\caption{The Hubble rate of the model I (flat $\Lambda$CDM represent the case
when $\alpha=0$ and $\omega=-1$) versus the redshift $z$. The points with bars indicate
the experimental data summarized in Table 1 of Ref \cite{ref36}. It is clear that Model I is best-fitted to data when we use joint
analysis of two datasets.}
\end{minipage}
\end{figure}
\begin{figure}[htb]
\centering
\includegraphics[width=15cm,height=15cm,angle=0]{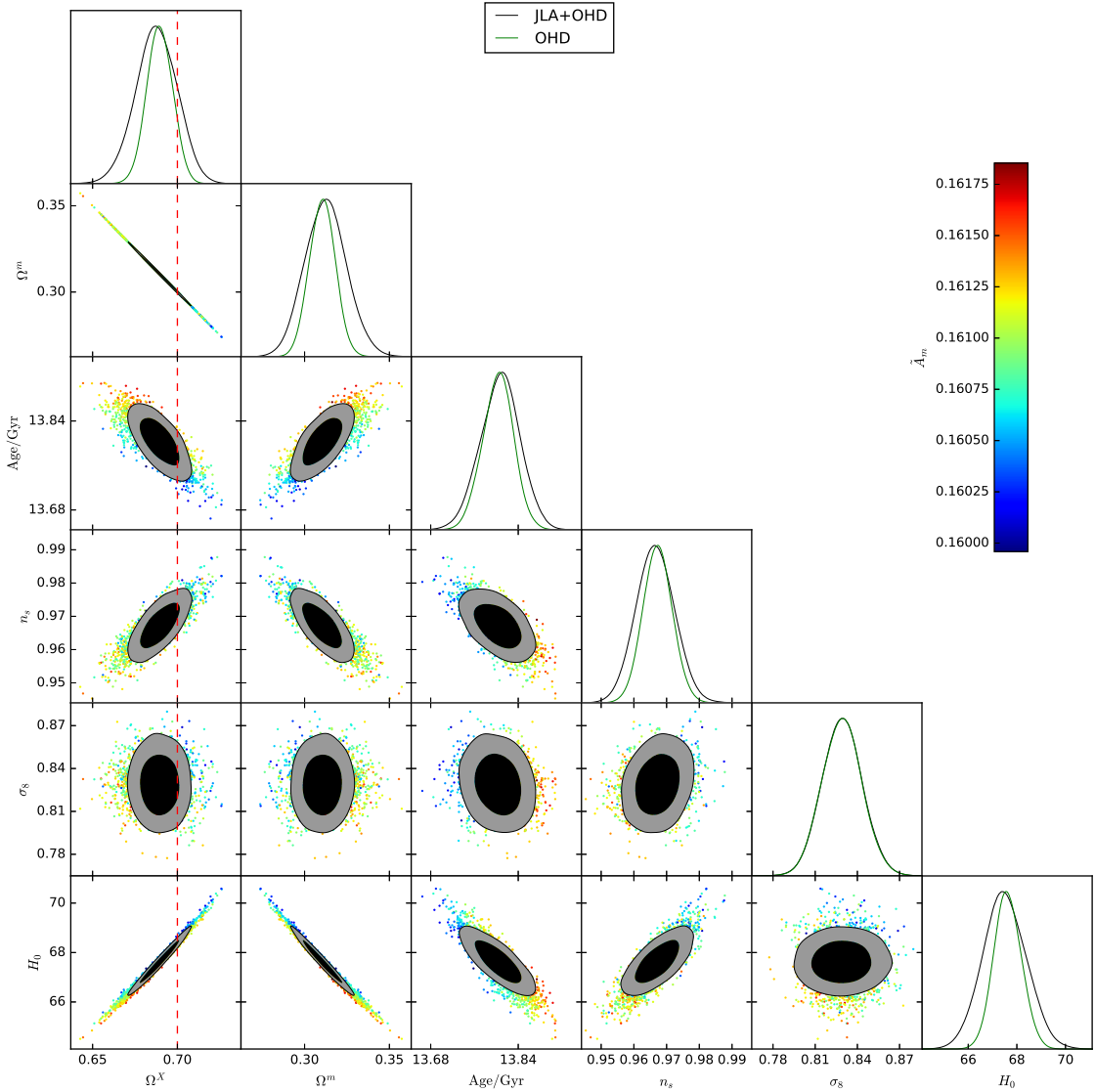}
\caption{$1\sigma$ and $2\sigma$ confidence regions from the fits of the $\omega$BI Model to the single OHD and the joint OHD+JLA analysis.
The vertical dotted Red line stands for $\Omega^{X}=0.7$.}
\end{figure}
\subsection{CONSTRAINTS ON $\omega$BI MODEL}
\begin{table}[ht]
\caption{Results from the fits of the $\omega$BI model to the data at 1$\sigma$ confidence level.}
\centering
\begin{tabular} {ccccccccc}
\hline
Parameter    & JLA & OHD & JLA+OHD  \\[0.5ex]           
\hline
\hline{\smallskip}
$H_{0}$ & -&   $67.94 \pm 1.6$ & $67.53\pm 1.1$\\              

$\Omega^{X}$ &  $0.624\pm 0.031$  &   $0.6920\pm 0.0052$ &  $0.6862\pm 0.0092$ &\\        

$\Omega^{m}$ &  $0.376\pm 0.031$ & $0.310\pm 0.0052$ & $0.3120\pm 0.0092$\\  

$n_{s}$ & $0.959\pm 0.019$ & $0.9663\pm 0.0023$ & $0.9648\pm 0.0046$\\

$\sigma_{8}$ & $0.826\pm 0.021$ & $0.825\pm 0.018$ & $0.823\pm 0.052$\\

$Age/Gyr$ & $14.018^{+0.035}_{-0.075}$ & $13.812\pm 0.046$ & $13.815\pm 0.031$\\

$A_{m}$ & $0.008\pm 0.005$ & $0.006\pm 0.007$ & $0.002\pm 0.004$\\

$\omega^{X}$ & $-0.91 \pm 0.65$  &  $-0.96 \pm 0.83$ & $-0.97 \pm 0.35$ \\[0.5ex]
\hline
\end{tabular}
\label{table:nonlin}
\end{table}
The $\omega$BI model considers two more free parameters, a DE fluid with a free EoS parameter $\omega$ with $-1<\omega<-\frac{1}{3}$
which we assume to be constant with respect to the cosmic expansion and a constant parameter $\alpha$ which is responsible for the
inherent anisotropy of the BI spacetime. Table 2 shows the results of our statistical analysis for Model II. Figure 4 depicts 
$1\sigma$ and $2\sigma$ confidence regions from the fits of the $\omega$BI model to the {\bf OHD} and {\bf OHD+JLA} datasets.
Table 2 shows that the matter density $\Omega^{m}$ derived from Model II by using OHD and the joint analysis (OHD+JLA) is again
in high agreement with the value obtained from Planck 2015 Collaboration. Also, although the estimates of the current expansion rate $H_{0}$
derived from OHD alone and the joint OHD+JLA analysis are in excellent consistent with results of Refs \cite{ref45,ref46} but still
are different from the results of Efstathiou \cite{ref47} and Riess et al \cite{ref48} by $2.1\sigma$ and $2.9\sigma$, respectively.
The robustness of our fits can be viewed by looking at figures 5 \& 6.\\
\begin{figure}[htb]
\begin{minipage}[b]{0.5\linewidth}
\centering
\includegraphics[width=8cm,height=6cm,angle=0]{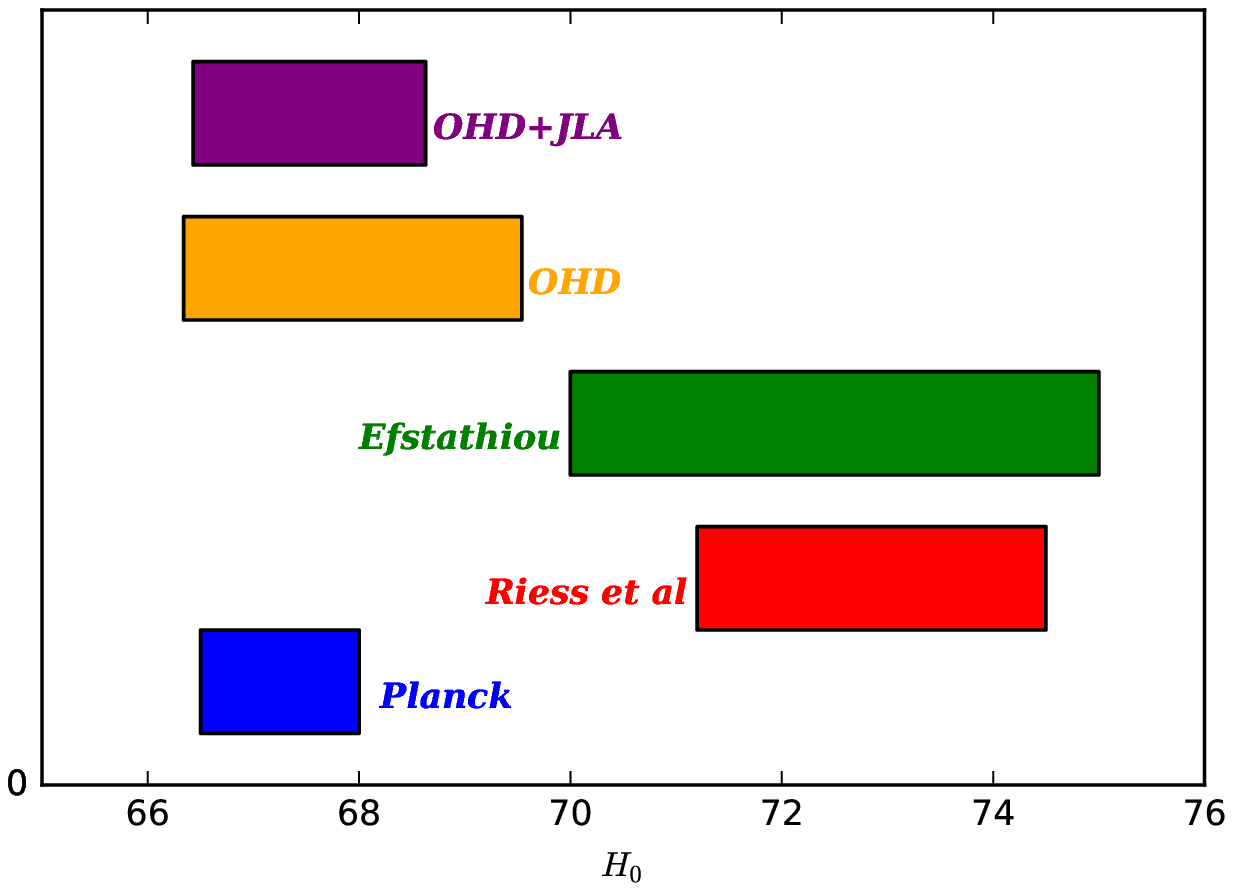} \\
\caption{A schematic representation at $1\sigma$ confidence region of $H_{0}$ for the Model II when fitted to OHD (gold color) and 
the joint OHD+JLA (purple). Constraints from other measurements have also been shown.}
\end{minipage}
\hspace{0.5cm}
\begin{minipage}[b]{0.5\linewidth}
\centering
\includegraphics[width=8cm,height=6cm,angle=0]{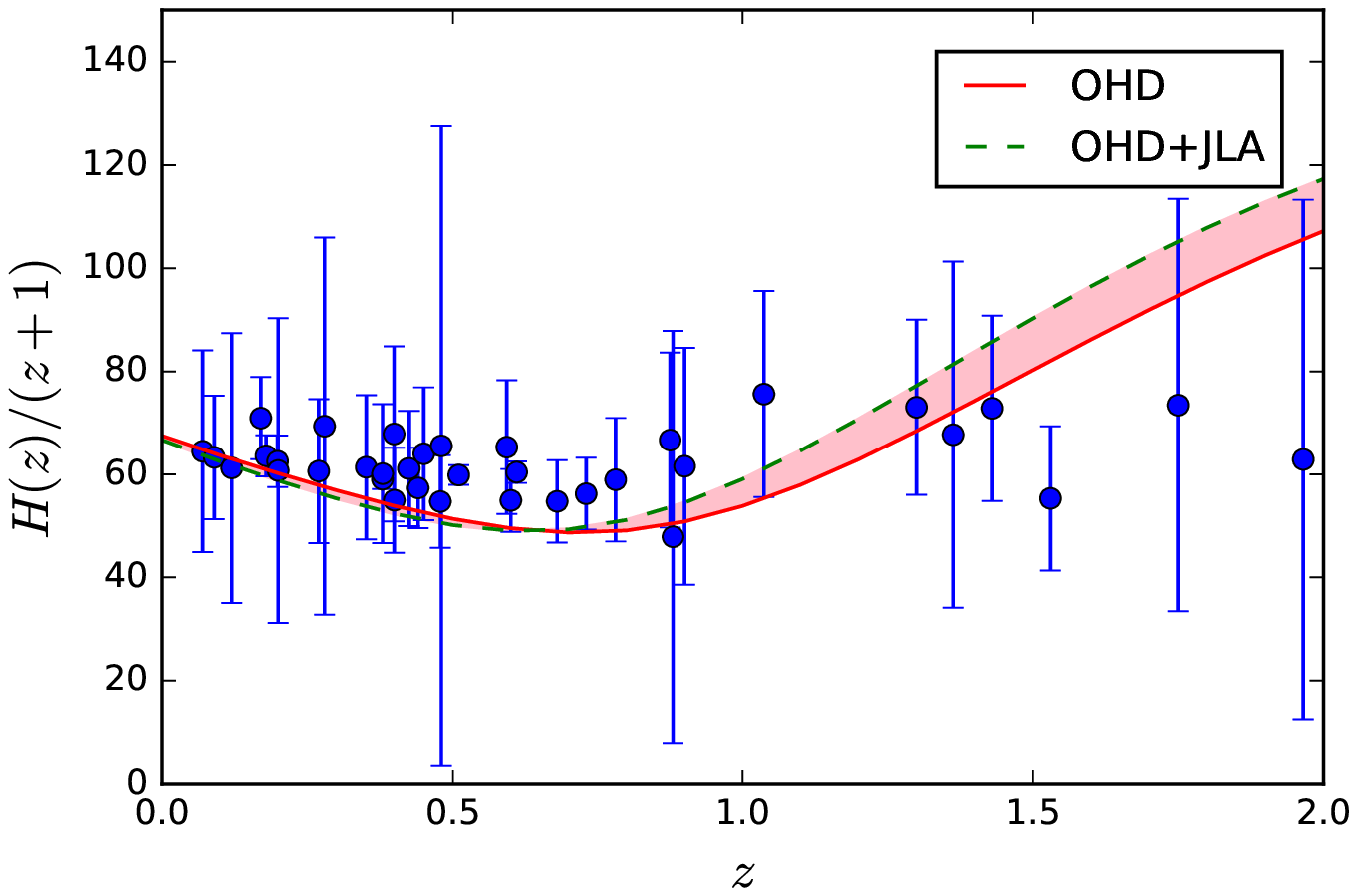}
\caption{The Hubble rate of the model II ($\omega$BI) represent the case
when $\alpha\neq0$ and $-1<\omega<-\frac{1}{3}$) versus the redshift $z$. This Model is best-fitted to data when we use joint
analysis of two datasets.}
\end{minipage}
\end{figure}
In what follows we use the Akaike information criterion (AIC) and Bayes factor ($\Psi$) in order to compare the considered cosmologies.
For a given theoretical model and a given dataset, the Akaike estimate of minimum information (Akaike 1974) \cite{ref49} is defined as
\begin{equation}
\label{eq12} AIC = -2\log L^{max} + 2N,
\end{equation}
\begin{table}[ht]
\caption{Comparison of the cosmological models by $\Delta(AIC)$ and $\Psi$ using joint OHD+JLA dataset.}
\centering
\begin{tabular} {ccc}
\hline
Model    & $\Delta(AIC)$ & $\Psi$  \\[0.5ex]           
\hline
\hline{\smallskip}
$\Lambda CDM$ & 0 &   $1$ \\              

$\omega BI$ &  $1.27$ & $1:1.18$\\[0.5ex]        

\hline
\end{tabular}
\label{table:nonlin}
\end{table}
where $N$ is the number of independent parameters of the Model. In this method, the preference is given to the model with the lowest AIC.
Similarly, we use the Bayes factor $\Psi$ which provides a criterion for choosing between two models by comparing their best likelihood values.
It is worth noting that $0\leq\Delta(AIC)\leq 2$ gives substantial evidence for the model, $\Delta(AIC)\leq 7$ shows
less support for the model and $\Delta(AIC)>10$ indicates that the model is unlikely.
Here we are interested in comparing Model II ($\omega$BI) with Model I ($\Lambda$CDM). Moreover, the Bayes
factor, $\Psi=L^{max}_{\Lambda CDM}/L^{max}_{\omega BI}$ represents the odds for the $\Lambda$CDM model against the
$\omega$BI model. We note that, odds lower than $1:10$ indicate a strong evidence against $\Lambda$CDM model
whereas odds greater than $10:1$ indicate a strong evidence against $\omega$BI model (Jeffreys \cite{ref50}).
The difference, $\Delta(AIC)=AIC_{\omega BI}-AIC_{\Lambda CDM}$, of the AIC value and the Bayes factor $\Psi$ of
$\omega$BI scenario against $\Lambda$CDM have been shown in Table 3. The value of $-2 \log L$ for $\omega$BI and $\Lambda$CDM models when
we fitted them to the OHD data alone, are found as $27.35$ and $28.03$ respectively which clearly shows
that $\omega$BI can be used to fit OHD data, this result is in agreement with what obtained by Wang \& Zhang \cite{ref51}. Also,
fitting Both models to JLA data alone, we obtain $-2 \log L_{\omega BI}=-212.53$ and $-2\log L_{\Lambda CDM}=-213.96$ which shows that
$\omega$BI can be used to fit the SN Ia data as well. The differences, $\Delta(AIC)$, of the AIC values
as well as the Bayes factor $\Psi$ are presented in Table 3.
\subsection{COSMOLOGICAL DECELERATION-ACCELERATION TRANSITION REDSHIFT}
It is well known that the expansion phase of universe changes from decelerating to accelerating at a specific redshift
called \textgravedbl transition redshift\textacutedbl $z_{t}$. Here, For completeness of our study, we derived $z_{t}$ for both models by using OHD
and combined OHD+JLA datasets. In general, the deceleration parameter is given by
\begin{equation}
\label{eq13} q(z)=-\frac{1}{H^{2}}\left(\frac{\ddot{a}}{a}\right)=\frac{(1+z)}{H(z)}\frac{dH(z)}{dz}-1.
\end{equation}
It is clear that the transition redshift is implicitly defined by the condition $q(z_{t})=\ddot{a}(z_{t})=0$. From (\ref{eq7}) 
one could easily find the transition redshift for Model I as
\begin{equation}
\label{eq14} z_{t}= \left(\frac{2\Omega^{\Lambda}}{\Omega^{m}}\right)^{\frac{1}{3}}-1.
\end{equation}
Also, from (\ref{eq7}), the transition redshift for $\omega$BI Model could be found as 
\begin{equation}
\label{eq15} z_{t}=\left(\frac{\Omega^{m}}{(\Omega^{m}-1)(1+3\omega^{X})}\right)^{\frac{1}{3\omega^{X}}}-1.
\end{equation}
\begin{table}[ht]
\caption{Deceleration parameters and Deceleration-Acceleration transition redshifts.}
\centering
\begin{tabular} {ccc}
\hline
Model    & $z_{t}(OHD)$ & $z_{t}(OHD+JLA)$  \\[0.5ex]           
\hline
\hline{\smallskip}
$\Lambda CDM$ & $0.642\pm 0.0155$ &   $0.637\pm 0.0244$ \\              

$\omega BI$ &  $0.671\pm 0.0185$ & $0.639\pm 0.0283$\\[0.5ex]        
\hline
\hline
$q_{\Lambda CDM}$ & $-5.43^{+0.38}_{-0.59}$ & $-5.56^{+0.49}_{-0.35}$ \\ 

$q_{\omega BI}$ & $-5.79^{+0.37}_{-0.45}$ & $-6.10^{+0.67}_{-0.21}$ \\[0.5ex] 
\hline
\end{tabular}
\label{table:nonlin}
\end{table}
Using the best-fit parameters given in tables 2 \& 3, we can obtain the transition redshift for $\Lambda$CDM and $\omega$BI models. 
The deceleration parameter as well as the deceleration-acceleration transition redshift with corresponding standard deviation
for each model have been shown in Table 4. From this table we observe that, in both Models, combining two datasets give rises
to more stringent constraints on transition redshift. These results show that the $\omega$BI model enters to the accelerating phase at earlier
time with respect to the $\Lambda$CDM model. But in both models, when we use joint OHD+JLA analysis to constrain model parameters, the transition
to accelerating phase is occur at lower redshifts. Figures 7\&8 depict the variation of deceleration parameter $q$ versus
redshift $z$ for both Models and different datasets.
\begin{figure}[htb]
\begin{minipage}[b]{0.5\linewidth}
\centering
\includegraphics[width=8cm,height=6cm,angle=0]{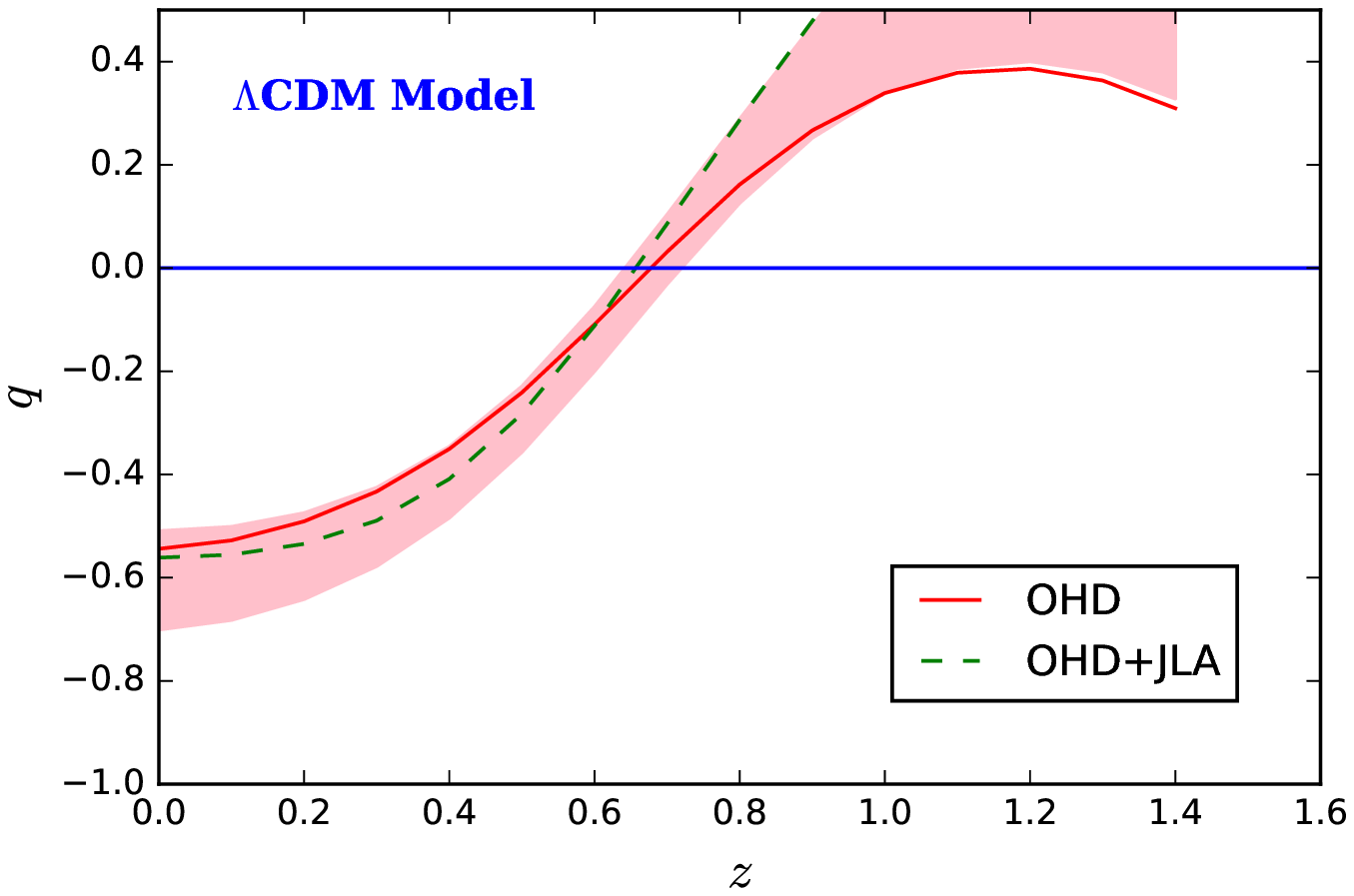} \\
\caption{variation of deceleration parameter $q$ versus redshift $z$ for $\Lambda CDM$ model by using OHD and OHD+JLA datasets. When we constrain
Model by joint OHD+JLA analysis, it enters to the accelerating phase at lower redshift.}
\end{minipage}
\hspace{0.5cm}
\begin{minipage}[b]{0.5\linewidth}
\centering
\includegraphics[width=8cm,height=6cm,angle=0]{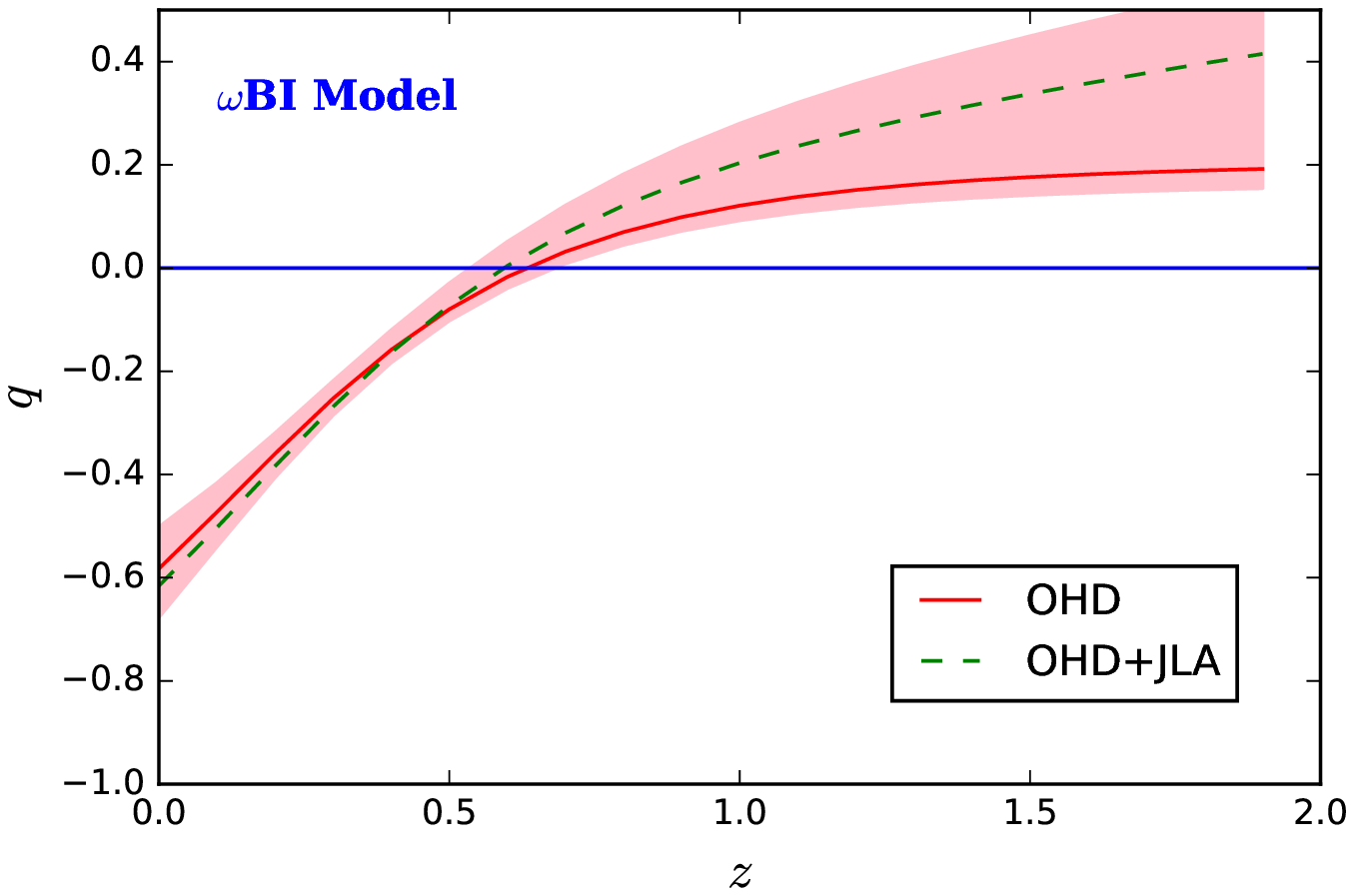}
\caption{variation of deceleration parameter $q$ versus redshift $z$ for $\omega$BI model by using OHD and OHD+JLA datasets. When we constrain
Model by joint OHD+JLA analysis, it enters to the accelerating phase at lower redshift.}
\end{minipage}
\end{figure}
\section{CONCLUDING REMARKS}
Instead of maximally symmetric FRW spacetime we considered Bianchi type I metric which is homogeneous but anisotropic. Based on recent cosmic
observations, this metric is better consistent with the line-element of real universe. Two datasets, Namely observational Hubble data (OHD) and
\textgravedbl latest joint light curves\textacutedbl (JLA) and their joint combination have been considered to constrain parameters of two
theoretical $\Lambda$CDM and $\omega$BI  models. In general, we found that using joint OHD+JLA dataset puts tighter constraints on the 
model parameters. Specially, by using joint OHD+JLA, the expansion rate of universe is obtained as $67.9 \pm 1.2$ and
$67.53 \pm 1.1$ for $\Lambda$CDM and $\omega$BI models respectively. These results are obviously in excellent consistent with results obtained
by cosmic microwave background (CMB) anisotropies \cite{ref45,ref52}, and baryon acoustic oscillation (BAO) \cite{ref46} projects.
Other results, specially current value of deceleration parameter $q_{0}$ and its corresponding value of the transition redshift $z_{t}$
for both models are found to be in good agreement with almost all recent cosmological observations.
\section*{ACKNOWLEDGMENT}
Author thanks The Mahshahr branch, Islamic Azad University for providing facility and support where this
work was carried out. The author is also grateful to the anonymous referee for valuable comments and suggestions.

\section*{APPENDIX A}
It have been shown in Refs \cite{ref38,ref39} that if we solve the Einstein's field equations (\ref{eq2}) for line-element (\ref{eq1}), the metric components could be obtained as
\setcounter{equation}{0}
\renewcommand{\theequation}{A\arabic{equation}}
\begin{equation}
A(t)=a_{1}a~ exp(b_{1}\int a^{-3}dt),
\end{equation}
\begin{equation}
B(t)=a_{2}a~ exp(b_{2}\int a^{-3}dt),
\end{equation}
\begin{equation}
C(t)=a_{3}a~ exp(b_{3}\int a^{-3}dt),
\end{equation}
where
\[
a_{1}a_{2}a_{3}=1,~~~~~~~b_{1}+b_{2}+b_{3}=0.
\]
The anisotropy parameter $A_{m}$ and the average Hubble parameter $H$ are defined as
\begin{equation}
A_{m}=\frac{1}{3}\left[\left(\frac{H_{x}-H}{H}\right)^{2}+\left(\frac{H_{y}-H}{H}\right)^{2}
+\left(\frac{H_{z}-H}{H}\right)^{2}\right],
\end{equation}
and
\begin{equation}
H=\frac{1}{3}(H_{x}+H_{y}+H_{z}),
\end{equation}
where directional Hubble parameters $H_{x},H_{y},$ and $H_{z}$ could be obtained as bellow
\begin{equation}
H_{x}=\frac{\dot{A}(t)}{A(t)}=(1-3b_{1}a^{-3})\frac{\dot{a}}{a},
\end{equation}
\begin{equation}
H_{y}=\frac{\dot{B}(t)}{B(t)}=(1-3b_{2}a^{-3})\frac{\dot{a}}{a},
\end{equation}
\begin{equation}
H_{z}=\frac{\dot{C}(t)}{C(t)}=(1-3b_{3}a^{-3})\frac{\dot{a}}{a}.
\end{equation}
Using eqs ($A6-A8$) and ($A5$) in eq ($A4$), we can find the expression of anisotropy parameter as
\begin{equation}
A_{m}=3a^{-6}(b_{1}^{2}+b_{2}^{2}+b_{3}^{2}).
\end{equation}
Since $K=b_{1}b_{2}+b_{1}b_{3}+b_{2}b_{3}$ (see ref\cite{ref39}), above equation could be written in the following compact form
\begin{equation}
A_{m}=-6Ka^{-6}=A_{m0}(1+z)^{6}.
\end{equation}
Note that, $K$ is a negative parameter. Equation ($A10$) clearly shows that as universe expands, the anisotropy of Bianchi type I space-time decreases and ultimately dies out at redshift $z=-1$. The current value of anisotropy parameter is $A_{m}=-6K$ (See figure 9 for more details). It is worth noting that to get FRW cosmology (i.e $A(t)=B(t)=C(t))$, one has to consider $b_{1}=b_{2}=b_{3}=0$ which leads to zero anisotropy parameter (i.e $A_{m}=0$)
\begin{figure}[htb]
	\centering
	\includegraphics[width=8cm,height=6cm,angle=0]{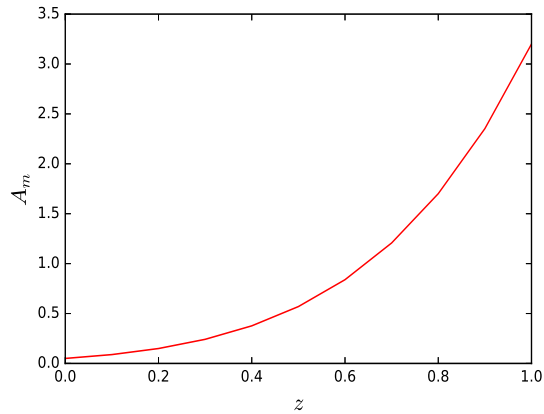}
	\caption{Variations of anisotropy parameter $A_{m}$ vs redshift $z$.}
\end{figure}


\begin{thebibliography}{99}

\bibitem {ref1}
S. Perlmutter, et al, Nature, {\bf 391}, 51 (1998).
\bibitem {ref2}
S. Perlmutter, et al, Astrophys. J, {\bf 517} 565 (1999).
\bibitem {ref3}
A. G. Riess, et al, Astron. J, {\bf 116}, 1009 (1998).
\bibitem {ref4}
M. Tegmark, M. Strauss, M. Blanton, et al, Phys. Rev. D, {\bf 69}, 103501 (2004).
\bibitem {ref5}
B. P. Schmidt, et al, Astrophys. J. {\bf 507}, 46 (1998).
\bibitem {ref6}
S. Weinberg, Rev. Mod. Phys, {\bf 61}, 1 (1989).
\bibitem {ref7}
P. J. E. Peebles, B. Ratra, Rev. Mod. Phys, {\bf 75}, 559 (2003).
\bibitem {ref8}
P. J. Steinhardt, L. Wang, and I. Zlatev, Phys. Rev. D {\bf 59}, 123504, (1999).
\bibitem {ref9}
M. Malquarti, E. J. Copeland, and A. R. Liddle, Phys. Rev. D, {\bf 68}, 023512, (2003).
\bibitem {ref10}
L. Amendola, Phys. Rev. D, {\bf 62}, 043511, (2000).
\bibitem {ref11}
A. Sandage, Physics Today, {\bf February 23}, 34, (1970).
\bibitem {ref12}
E. Hubble, Proceedings of the National Academy of Science, {\bf 15}, 168 (1929).
\bibitem {ref13}
R. P. Kirshner, Proceedings of the National Academy of Science, {\bf 101}, 8, (2003).
\bibitem {ref14}
A. G. Riess et al, Astrophys. J, {\bf 826}, 56, (2016).
\bibitem {ref15}
G. Hinshaw, et al, ApJ, {\bf 208}, 19, (2013).
\bibitem {ref16}
Planck Collaboration, A\&A {\bf 594}, A13 (2016). 
\bibitem {ref17}
D. J. Fixsen, Astrophys. J, {\bf 707}, 916, (2009).
\bibitem {ref18}
J. L. Sievers, et al, J. Cosmol. Astropart. Phys, {\bf 1310}, 060, (2013).
\bibitem {ref19}
J. R. Gott, M. S. Vogeley, S. Podariu and B. Ratra, Astrophys. J, {\bf 549}, 1, (2001).
\bibitem {ref20}
S. Podariu, et al, ApJ, {\bf 559}, 9, (2001).
\bibitem {ref21}
G. Chen and B. Ratra, PASP, {\bf 115}, 1143, (2003).
\bibitem {ref22}
S. Crandall and B. Ratra, ApJ, {\bf 815}, 87, (2015).
\bibitem {ref23}
S. Crandall and B. Ratra, Phys. Lett. B, {\bf 732}, 330, (2014).
\bibitem {ref24}
X. Ding, et al, ApJ, {\bf 803}, L22, (2015).
\bibitem {ref25}
O. Farooq and B. Ratra, Astrophys. J, {\bf 766}, L7, (2013). 
\bibitem {ref26}
G. Chen and B. Ratra, PASP, {\bf 123}, 1127, (2011).
\bibitem {ref27}
Y. Chen and B. Ratra, Phys. Lett. B, {\bf 703}, 406, (2011).
\bibitem {ref28}
R. Jimenez and A. Loeb, ApJ, {\bf 573}, 37, (2002).
\bibitem {ref29}
Z.-E, Liu, et al,  Phys. Lett. B, {\bf 14}, 21, (2014).
\bibitem {ref30}
V. V.Lukovic, R. D\textasciiacute Agostino and N. Vittorio, A \& A, {\bf 595}, A109, (2016).
\bibitem {ref31}
J. P. Zibin, Phys. Rev. D, {\bf 78}, 043504, (2008).
\bibitem {ref32}
J. P. Zibin, Phys. Rev. D, {\bf 84}, 123508, (2011).
\bibitem {ref33}
W. Valkenburg, V. Marra and C. Clarkson, Mon. Not. R. Astron. Soc, {\bf 438}, L6, (2012).
\bibitem {ref34}
M. Zumalacarregui, J. Garcia-Bellido and P. Ruiz-Lapuente, JCAP, {\bf1210}, 009, (2012).
\bibitem {ref35}
M. Tokutake, K. Ichiki and C-M. Yoo, arXiv: 1712.04229 (2017).   
\bibitem {ref36}
O, Farooq, F. R. Madiyar, S. Crandall and B. Ratra, Astrophys. J, {\bf 835}, (2017).
\bibitem {ref37}
M. Goliath, G. F. R. Ellis, Phys. Rev D, {\bf 60}, 023502, (1999).
\bibitem {ref38}
B. Saha, Mod. Phys. Lett. A, {\bf 20}, 2127, (2005).
\bibitem {ref39}
H. Amirhashchi, Astrophys. Space Sci, {\bf 351}, 641, (2014).
\bibitem {ref40}
M. Betoule, et al, A\&A, {\bf 568}, A22, (2014).
\bibitem {ref41}
J. Tr\o{}st Nielsen, A. Guffanti, and S. Sarkar, Scientific Reports, {\bf 6}, 35596, (2016).[arXiv:1506.01354]
\bibitem {ref42}
D. Blas, J. Lesgourgues and T. Tram, JCAP, {\bf 1107}, 034, (2011).
\bibitem {ref43}
B. Audren, J. Lesgourgues, K. Benabed and S. Prunet, JCAP, {\bf 1302}, 001 (2013).
\bibitem {ref44}
https://github.com/cmbant/getdist
\bibitem {ref45}
P. A. R. Adel, et al, A\& A, {\bf 594}, A13, (2016). 
\bibitem {ref46}
E. Aubourg, et al, Phys. Rev D, {\bf 92}, 123516, (2015).
\bibitem {ref47}
G. Efstathiou, MNRAS, {\bf 440}, 1138, (2014).
\bibitem {ref48}
A. G. Riess, et al, APJ, {\bf 826}, 1, (2016).
\bibitem {ref49}
H. Akaike, IEEE Transactions on Automatic Control, {\bf 19}, 716, (1974).
\bibitem {ref50}
H. S. Jeffreys, Theory of probability, The International series of monographs on physics, Oxford: Clarendon Press New York, (1983).
\bibitem {ref51}
H. Wang and T.-J. Zhang, ApJ, {\bf 748}, 111, (2012).
\bibitem {ref52}
D. Spergel, et al, Astrophys. J. Suppl, {\bf 148}, 170, (2003).
\end{thebibliography}
\end{document}